\documentstyle[psfig]{mn}

\begin{document}
\title[Far-IR Quasars]{Far Infrared Loud Quasars 1: Disturbed and Quiescent
Quasars in the PG Survey}
\author[D.L. Clements]
{D.L. Clements$^{1,2}$\\ $^1$Department of Physics and Astronomy,
University of Wales Cardiff, PO Box 913, Cardiff, CF24 3YB\\
$^2$Institut d'Astrophysique Spatiale, B\^atiment 121, Universite
Paris XI, F-91405 ORSAY CEDEX, France\\ }

\maketitle

\begin{abstract}
We use host galaxy imaging studies of the PG Quasar survey to compare
the far-infrared (FIR) properties of quasars with disturbed and undisturbed host
galaxies. By using survival
analysis, we show that the quasars with disturbed host galaxies, with
morphologies classified from a homogenous data set, have a 60$\mu$m
luminosity distribution that is different from that of those with
undisturbed hosts with $>$97\% confidence. For morphological classifications
using an inhomogenous data set, including HST data for some objects,
this confidence rises to $>$99\% confidence.
The mean 60$\mu$m luminosity of
the disturbed--host quasars is several times greater than that of
the undisturbed-host quasars. However,
possible biases in the PG survey might affect
these conclusions. Our results are interpreted as supporting the idea
that quasars are related to at least some Ultraluminous Infrared Galaxies. We
discuss the implications of this result for studies of quasar and
galaxy evolution.
\end{abstract}


\section{Introduction}

The host galaxies of quasars have been the subject of intense
observational scrutiny in recent years. A general picture had
emerged that radio-loud quasars lie in giant ellipticals, whilst
radio-quiet quasars are in spirals (e.g. Hutchings et al.
\shortcite{h89}).  Recent results, however, have confused this
picture, with some radio-quiet quasars showing signs of interacting or
elliptical hosts. The
most recent HST data (McLure et al. 1998) goes further to suggest that
essentially all bright quasars (M$_R < -24$), whether radio loud or
radio quiet, lie in elliptical galaxies. Meanwhile, there is mounting
evidence that interactions and mergers between galaxies are involved
in triggering QSO-like activity (eg. Stockton
\shortcite{s90}, Hutchings \& Neff
\shortcite{h92}).  There are also a number of suggestions that some
Ultraluminous Infrared Galaxies (ULIRGs) are an early stage in the
evolution of quasars (Sanders \& Mirabel \shortcite{s96} and
references therein), though much of the
luminosity in the ULIRG-phase appears to be starburst rather than AGN powered {\cite{g98}.
ULIRG activity is strongly linked to galaxy
interactions and mergers \cite{l94}. Clements et al.
\shortcite{c96a}, for
example, showed that 90\% of a sample of 60 ULIRGs are in disturbed
systems, a result confirmed by Murphy et al. (1996) for a similar but
complimentary ULIRG sample.

These results lead to a possible scenario for quasar triggering in
which a galaxy merger causes gas to collapse into the nucleus of a
galaxy (eg. Mirabel \shortcite{m93}), initiating a starburst and
fuelling an AGN. The resulting emission is obscured by dust leading to
a ULIRG-like phase. The central engine then expels or destroys the
obscuring material until it can be observed directly, as is the case
with broad line ULIRGs (eg. Mrk231 and Mrk1014). The merging
galaxies eventually settle into a stable morphology, and signs of the
starburst will fade. The details of this scheme depend on the
relative lifetimes of the ULIRG and quasar stages.  In this scenario
quasars in disturbed or interacting
hosts should be at an earlier stage in their evolution from ULIRG-like
object to conventional quasar. The defining characteristic of a
ULIRG is extreme FIR luminosity, L$_{FIR}>10^{12}L_{\odot}$. Therefore,
the ULIRG-to-quasar evolutionary scenario predicts that quasars
in disturbed host galaxies are more likely to have high FIR
luminosities than quasars in undisturbed hosts.

ISO observations, of known quasars (eg. Wilkes et
al. \shortcite{w97}), and large area surveys which discover new FIR-bright
quasars (eg. Oliver et al. \shortcite{o97}), will provide
significant new insights into these issues. Before these measurements
are completed, though, it is timely to re-examine
the properties of FIR-bright quasars as determined by the IRAS
satellite and ground-based observations. Elsewhere (Clements et al., in
preparation) we will discuss the results of imaging observations of a
sample of quasars selected for high FIR luminosity. Here we take the
alternative approach and compare the FIR properties of quasars in
disturbed and undisturbed hosts to see if there is indeed a
connection between disturbed host morphology and high FIR luminosity.

The rest of the paper is organised as follows. In the next section we
describe the quasar sample and imaging data used for this study and
arrive at morphological classifications for the host galaxies. In
Section 3 we discuss the statistical methods used to compare the FIR
properties of the disturbed and undisturbed quasars and find a
significant difference between the two classes.
These results are discussed in Section 4
and we summarise our conclusions in Section 5.

\begin{table}
\label{qsos}
\begin{tabular}{lrrrr}
Object&Redshift&F$_{60}$&F$_{100}$&L$_{FIR}$\\ \hline
0026+129  &0.142    &$<$27    &$<$80    &$<$8.1\\
0052+251  &0.155	&93	&$<$338	&$<$37\\
0157+001  &0.164	&2377	&2322	&610\\
0923+201  &0.190	&$<$300	&$<$1000	&$<$170\\
0947+396  &0.206	&201	&462	&110	\\
0953+414  &0.239	& $<$129	&$<$315	&$<$100\\
1012+008  &0.185	&$<$140	&$<$347	&$<$66\\
1048+342  &0.167	&$<$140	&$<$347	&$<$53\\
1121+422  &0.224	&$<$140	&$<$315	&$<$94 \\
1151+117  &0.176	&$<$154	&$<$347	&$<$62 \\
1202+281  &0.165	&110	&420	&52 \\
1307+085  &0.155	&$<$154	&$<$347	&$<$48\\
1309+355  &0.184	&$<$140	&$<$347	&$<$65 \\
1322+659  &0.168	&$<$89	&$<$257	&$<$37 \\
1352+183  &0.158	&$<$140	&$<$347	&$<$47 \\
1354+213  &0.300	&$<$154	&$<$347	&$<$190 \\
1402+261  &0.164	&229	&340	&67  \\
1427+480  &0.221	&$<$112	&$<$252	&$<$73 \\
1444+407  &0.267	&117	&170	&94  \\
1613+658  &0.129	&635	&1090	&120	\\
1700+518  &0.292	& 480    & 482  &420\\
0050+124  &0.061	&2293	&2959	&85	\\
0804+761  &0.100	&191	&$<$315	&$<$21 \\
0838+770  &0.131	&174	&426	&40  \\
0844+349  &0.064	& 163	&294	&7.5	\\
1001+054  &0.161	&27	&$<$69	&$<$9.7 \\
1114+445  &0.144	&191	&$<$347	&$<$47 \\
1115+407  &0.154	& $<$140	&$<$347	&$<$45 \\
1211+143  &0.085	&305	&689	&28  \\
1229+204  &0.064	&163	&$<$462	&$<$9.3 \\
1351+640  &0.087	& 757	&1184	&62	\\
1404+226  &0.098	&$<$154	&$<$347	&$<$19 \\
1411+442  &0.089	&162	&$<$175	&$<$12 \\
1415+451  &0.114	&112	&260	&19  \\
1416-129  &0.129	&$<$140	&$<$315	&$<$30 \\
1426+015  &0.086	&318	&$<$315	&$<$22 \\
1435-067  &0.129	&$<$126	&$<$315	&$<$28 \\
1440+356  &0.077	&652	&1061	&42  \\
1519+226  &0.137	&$<$112	&$<$252	&$<$27 \\
1552+085  &0.119	&$<$126	&$<$315	&$<$24 \\
1612+261  &0.131	&$<$54	&$<$161	&$<$14 \\
1617+175  &0.114	&$<$98	&$<$252	&$<$17 \\
1626+554  &0.133	&$<$70	&$<$238	&$<$20 \\
2130+099  &0.061	&479	&$<$1000	&$<$21 \\
2214+139  &0.067	&337	&$<$282	&$<$13 \\
\end{tabular}
\caption{Properties of PG Quasars used for this study.}

F$_{60}$ and F$_{100}$ are the 60 and 100 $\mu$m fluxes in mJy,
respectively, L$_{FIR}$ is the FIR luminosity in units of
10$^{10}L_{\odot}$.
\end{table}

\section{Quasar Host Imaging Samples}

To assess any difference in the properties of
quasars in disturbed and undisturbed
hosts, we need a complete sample of objectively selected quasars for
which host galaxy imaging is available. Furthermore, since we are
interested in the FIR properties of these objects, we also need a
sample where complete FIR observations are available.

There are very few quasar samples for which complete host-imaging data
is available. Most quasar host studies to date have concentrated on
small samples of quasars with a range of properties (eg. Bahcall et al.
\shortcite{b96}, Boyce et al. \shortcite{b98}, McLure et al. \shortcite{m98}),
or have been selected from some particularly special
sub-class of quasar (eg. Boyce et al. \shortcite{b97}).  The only
survey which fits our requirement for a quasar sample with
complete imaging observations is the low redshift (z$<$0.3) PG quasar survey
(see Table 1).
Extensive imaging data for these objects is available, including
H-band observations by McLeod \&
Rieke (1994a,b) for all the objects, HST observations for many of the
high luminosity objects (\cite{ba97}, \cite{m98}), and extensive ground-based
imaging studies at both optical and infrared wavelengths (see Table 2 and 3).
These cover a well defined subsample of the PG quasar
survey \cite{s83}, including all objects with redshift z$<$ 0.3 and B
band absolute magnitude $M_B<$-22.

We take FIR data on these
objects from Sanders et al. \shortcite{s89}, which details IRAS
observations in both survey and pointed mode for the complete PG
sample. One of the objects in the parent sample, 1116+215, for which
IRAS data are not available, is removed from further consideration. We
also remove all four radio-loud quasars in the McLeod sample. Only one
of these, 3c273, was detected by IRAS, and this emission is strongly
dominated by non-thermal effects. Any non-thermal contribution to the
FIR emission of the radio quiet quasars in the final sample is below
the 10mJy level.

FIR and 60$\mu$m luminosities are calculated assuming H$_0$=75 kms${-1}$Mpc$^{-1}$and
q$_0$=0.5, and using the standard conversion from 60 and 100 $\mu$m
flux
\cite{h85}. The final sample, together with FIR luminosities and basic
data, is given in Table 1. The sample contains 45 radio--quiet
quasars, of which 14 are in disturbed systems.

Assessing the morphology of the host galaxies of these quasars is
critical to the present study and can be a difficult task. One may
take one of two general approaches to the present
data. We can choose to use homogeneous data, in this case the McLeod
\& Rieke H-band images, which may not be the best for the task but
which will be fairly uniform in the information they provide on an
individual object. Alternatively, one can take an inhomogeneous
approach and use as much existing data as possible for any given
object in the sample to arrive at what might be described as the best
classification for that object given the data available. There are
advantages and disadvantages to both approaches. In the homogeneous
approach, it is clear that the McLeod \& Rieke H-band images are not
as well--suited to host galaxy studies as, say, deep HST images, since
they are of limited angular resolution and depth. One is thus likely
to be more subject to errors in morphological classification. In
contrast, the inhomogeneous approach uses the best data available,
including HST images for the sample discussed here. However, such data is
not available for all objects, and so one might be subject to biases
from the selection of PG quasars observed by HST.

Since each method has its own benefits and problems, we will use each
in turn.

\subsection{Homogeneous Image Classifications}

The H-band observations of McLeod \& Rieke provide us with a
homogeneous imaging sample for the radio--quiet nearby PG quasars. To
determine their morphologies the images were independently examined by
three observers. They gave one of two classifications: class 0 for
quiescent, undisturbed quasar hosts, and class 1 for disturbed
hosts. In cases of disagreement the majority view is adopted. It would
be nice to attempt a classification based on strength of
merger. However, even with just two morphological classifications we
are already running into problems with small number statistics. Therefore,
for the present study, we must restrict ourselves to just the two
classes `disturbed' and `quiescent'. Our classifications are
shown in Table 2.

\subsection{Inhomogenous Image Classifications}

In Table 3
we summarise the results of host imaging studies of the objects in
this sample from ten different literature sources. Most of these
papers are concerned only with the imaging of host galaxies in general
and are not biased towards studies aimed at FIR-luminous objects. One
of the papers (Stockton et al. \shortcite{s98}) is aimed at a specific
object, PG1700+518, while only two, \cite{h92}, \cite{s88b} {\em are}
aimed at FIR luminous objects. These latter studies, though, only
contribute corroborative classifications for other studies that are
not biased in favour of FIR luminous objects. We thus hope to avoid
biases introduced by having more data available for FIR luminous than
for FIR faint objects.

For each object, a morphological class is provided by each study in
which it is observed: class 0 for quiescent, undisturbed quasar hosts,
and class 1 for disturbed hosts. A few objects have unclear
classifications in some studies, and are given a ? designation. Most
of these classes, excluding those based on the Mcleod \& Rieke images,
are provided by the authors of the collated papers. Classifications
based on the McLeod and Rieke data are taken from Table 2.

A final morphological assessment is arrived at by comparing the
results of all surveys that deal with a given object. Studies reaching
fainter magnitudes and especially those working at higher spatial
resolutions, using HST or adaptive optics, are given a higher
weighting in deciding the final assessment.

A number of objects have seemingly contradictory morphological
classifications in different studies. These are now discussed.

{\bf 0052+251}

Several bright objects near to the quasar might be thought to be
interaction companions based on ground-based observations (eg. McLeod
et al. \shortcite{m94b}, Dunlop et al. \shortcite{d93}, Hutchings et al.
\shortcite{h89}). However, HST images \cite{ba97} reveal that the
claimed second and third nuclei are in fact respectively an HII region
in the host galaxy's spiral arms and a foreground star. The host
appears to be an undisturbed spiral in the HST images. We therefore
classify this object as quiescent.

{\bf 0923+201}

This object is a member of a small group of galaxies, and on this
basis was thought to be interacting. HST data (Bahcall et
al. \shortcite{ba97}, McLure et al. \shortcite{m98}) however show no
signs of interaction, and the host galaxy appears to be a normal E2
elliptical. We thus classify this object as quiescent.

{\bf 0953+414}

The original H band images of McLeod \& Rieke \shortcite{m94b} show
few signs of disturbance, consistent with initial HST observations by
Bahcall et al. \shortcite{ba97}. However, deeper optical \cite{h89} and
IR \cite{d93} images from the ground suggested the presence of some
disturbance. This has now been confirmed by deeper HST images
\cite{m98} which reveal a tidal tail and up to four companions. We
thus classify this object as disturbed.

{\bf 1444+407}

Optical images \cite{h89} suggested that this object
might have an off-centre nucleus, indicating a disturbed morphology.
However, ground-based infrared \cite{m94b} and further optical
\cite{h92} observations have not confirmed this. HST images \cite{ba97}
show no signs of disturbance, but suggest the
presence of a nuclear bar. We thus classify this object as quiescent.

{\bf 1700+518}

H band imaging of this object by McLeod \& Rieke \shortcite{m94b}
shows no clear signs of disturbed structure. However, suggestions of
disturbance were seen in ground-based optical data \cite{h91}. The
situation has now been clarified by an adaptive optics study by
Stockton et al. \shortcite{s98} which has conclusively shown that this
quasar is interacting with a nearby companion. Both companion and
quasar host have tidal tails, and the redshift of the companion has
been confirmed to be the same as the quasar. We must thus conclude
that this is in fact an interacting system.

{\bf 1211+143}

The H band imaging from McLeod \& Rieke \shortcite{m94a} shows no signs of
disturbance. However, deeper K band imaging by Dunlop et al.
\shortcite{d93} reveals a bridge linking the quasar to a companion
galaxy. On this basis we classify this object as disturbed.

{\bf 1229+204}

This object appears undisturbed in the H band images from McLeod \& Rieke 
\shortcite{m94a}.
However, it is described as I? (possible interaction) by Hutchings et al.
\shortcite{h91}. Since the evidence for an interaction is only marginal, we
classify this object as quiescent.

{\bf 1351+640}

Optical observations have classified this object both as disturbed
\cite{h91} and quiescent \cite{h92}. The H band observations
\cite{m94a} suggest that this object
has an offset nucleus and asymmetric inner isophotes. We thus
tentatively classify this object as disturbed.

{\bf 1612+261}

This object was originally described as I?, a possible interaction
\cite{h89}. However, no signs of disturbance are visible in the H band
images \cite{m94a}. Since there is only marginal evidence for
disturbance, we classify this object as quiescent.

{\bf 2130+099}

Ground--based optical imaging \cite{h92}, \cite{h91} suggests that
this object is interacting. H band images \cite{m94a}, however, reveal
no signs of disturbance. In contrast to this, deeper K band imaging by
Dunlop et al. \shortcite{d93} show similar structures to the optical
-- bridges and companion objects -- and reveal an additional companion
not seen in the optical. We thus conclude that this object is
disturbed.

\begin{table}
\begin{tabular}{lrrrr}
Object&DLC&ACB&LD&Adopted\\ \hline
0026+129&0&0&1&0\\
0052+251&0&0&1&0\\
0157+001&1&1&1&1\\
0923+201&1&0&1&1\\
0947+396&1&1&1&1\\
0953+414&0&0&1&0\\
1012+008&1&1&1&1\\
1048+342&1&1&1&1\\
1121+422&0&0&0&0\\
1151+117&0&0&1&0\\
1202+281&1&1&1&1\\
1307+085&0&0&1&0\\
1309+355&0&1&0&0\\
1322+659&0&0&1&0\\
1352+183&0&0&0&0\\
1354+213&0&0&1&0\\
1402+261&0&0&0&0\\
1427+480&0&0&1&0\\
1444+407&0&0&0&0\\
1613+658&1&1&1&1\\
1700+518&0&0&0&0\\
0050+124&1&1&1&1\\
0804+761&0&0&0&0\\
0838+770&0&1&0&0\\
0844+349&1&1&1&1\\
1001+054&0&0&1&0\\
1114+445&0&1&0&0\\
1115+407&1&1&1&1\\
1211+143&0&1&0&0\\
1229+204&0&0&1&0\\
1351+640&1&0&0&0\\
1404+226&0&0&1&0\\
1411+442&0&0&1&0\\
1415+451&0&0&0&0\\
1416-129&0&0&1&0\\
1426+015&0&1&1&0\\
1435-067&0&0&1&0\\
1440+356&0&0&0&0\\
1519+226&0&0&0&0\\
1552+085&0&0&1&0\\
1612+261&0&0&1&0\\
1617+175&0&0&0&0\\
1626+554&0&0&0&0\\
2130+099&0&0&0&0\\
2214+139&0&1&0&0\\
\end{tabular}
\caption{Homogeneous morphological classifications for PG quasars used in this study.}
Objects classified as quiescent are indicated by 0, while 1 indicates
classification as a disturbed object.  Classifications were made
independently by three experienced observers, including the
author. The adopted morphology was then assigned the majority value in
cases of disagreement.
\end{table}

\begin{table*}
\begin{minipage}{80mm}
\begin{tabular}{lrrrrrrrrrrr}
Object&\multicolumn{10}{c}{Paper}&Adopted\\
&1&2&3&4&5&6&7&8&9&10&\\ \hline
0026+129&&0&&&&&&&&&0\\
0052+251&&0&&1&0&&&1&&&0\\
0157+001&&1&&1&&&1&&1&1&1\\
0923+201&&1&&0&0&&&1&&0&0\\
0947+396&&1&&&&&&&&&1\\
0953+414&&0&&1&?&&&1&&1&1\\
1012+008&&1&&1&1&&&1&&1&1\\
1048+342&&1&&&&&&&&&1\\
1121+422&&0&&&&&&&&&0\\
1151+117&&0&&&&&&&&&0\\
1202+281&&1&&&1&&&&&&1\\
1307+085&&0&&&0&&&&&&0\\
1309+355&&0&&&0&&&&&&0\\
1322+659&&0&&&&&&&&&0\\
1352+183&&0&&&&&&&&&0\\
1354+213&&0&&&&&&&&&0\\
1402+261&&0&&&0&&&&&&0\\
1427+480&&0&&&&&&&&&0\\
1444+407&&0&&&0&0&&1&&&0\\
1613+658&&1&&&&1&&&&&1\\
1700+518&&0&1&&&&1&&&&1\\
0050+124&1&&&&&&1&&1&&1\\
0804+761&0&&&&&&&&&&0\\
0838+770&0&&&&&&&&&&0\\
0844+349&1&&&&&&1&&&&1\\
1001+054&0&&&&&&&&&&0\\
1114+445&0&&&&&&&&&&0\\
1115+407&1&&&&&&&&&&1\\
1211+143&0&&&1&&&&&&&1\\
1229+204&0&&&&&?&&&&&0\\
1351+640&1&&&&&0&1&&&&1\\
1404+226&0&&&&&&&&&&0\\
1411+442&0&&&&&&&&&&0\\
1415+451&0&&&&&&&&&&0\\
1416-129&0&&&&&&&&&&0\\
1426+015&0&&&&&&&&&&0\\
1435-067&0&&&&&&&&&&0\\
1440+356&0&&&0&&&&&&&0\\
1519+226&0&&&&&&&&&&0\\
1552+085&0&&&&&&&&&&0\\
1612+261&0&&&&&&&1&&&0\\
1617+175&0&&&&&&&&&&0\\
1626+554&0&&&&&&&&&&0\\
2130+099&0&&&1&&1&1&&&&1\\
2214+139&0&&&&&0&&&&&0\\
\end{tabular}
\caption{Inhomogeneous morphological
classifications for PG quasars used in this study.}
Objects classified as quiescent in a given study are indicated by 0,
while 1 indicates classification as a disturbed object. An unclear
classification is indicated by ? for two objects.  Papers referred to
are: 1=\cite{m94a}; 2=\cite{m94b}; 3=\cite{s98}; 4=\cite{d93};
5=\cite{ba97}; 6=\cite{h92}; 7=\cite{h91}; 8=\cite{h89};
9=\cite{s88b}; 10=\cite{m98}
\end{minipage}
\end{table*}

\section{Comparison of Disturbed and Undisturbed Quasars}

Figure 1 compares the FIR luminosities of the disturbed and
undisturbed quasars as a function of redshift for both the homogeneous
and inhomogeneous classifications.  There is a clear tendency for the
disturbed quasars to have higher FIR luminosities. However, many of
the objects in this sample only have IRAS upper limits. Detection rate
statistics are thus also of interest. Of the 10 (15) quasars in
disturbed hosts in the homogeneous (inhomogeneous) set, 6 (11) are
detected at 60$\mu$m (i.e. 60\% (73\%)), whilst of the 35 (30)
undisturbed host quasars, 17 (12) are detected (i.e. 48\% (40\%)). At
100 $\mu$m, 6 (9) of the 10 (15) disturbed host quasars are detected
(60\% (60\%)) but only 8 (5) (22\% (16\%)) of the undisturbed host
quasars. This is only indicative since the FIR flux limits for this
sample are not uniform.

\begin{figure*}
\psfig{file=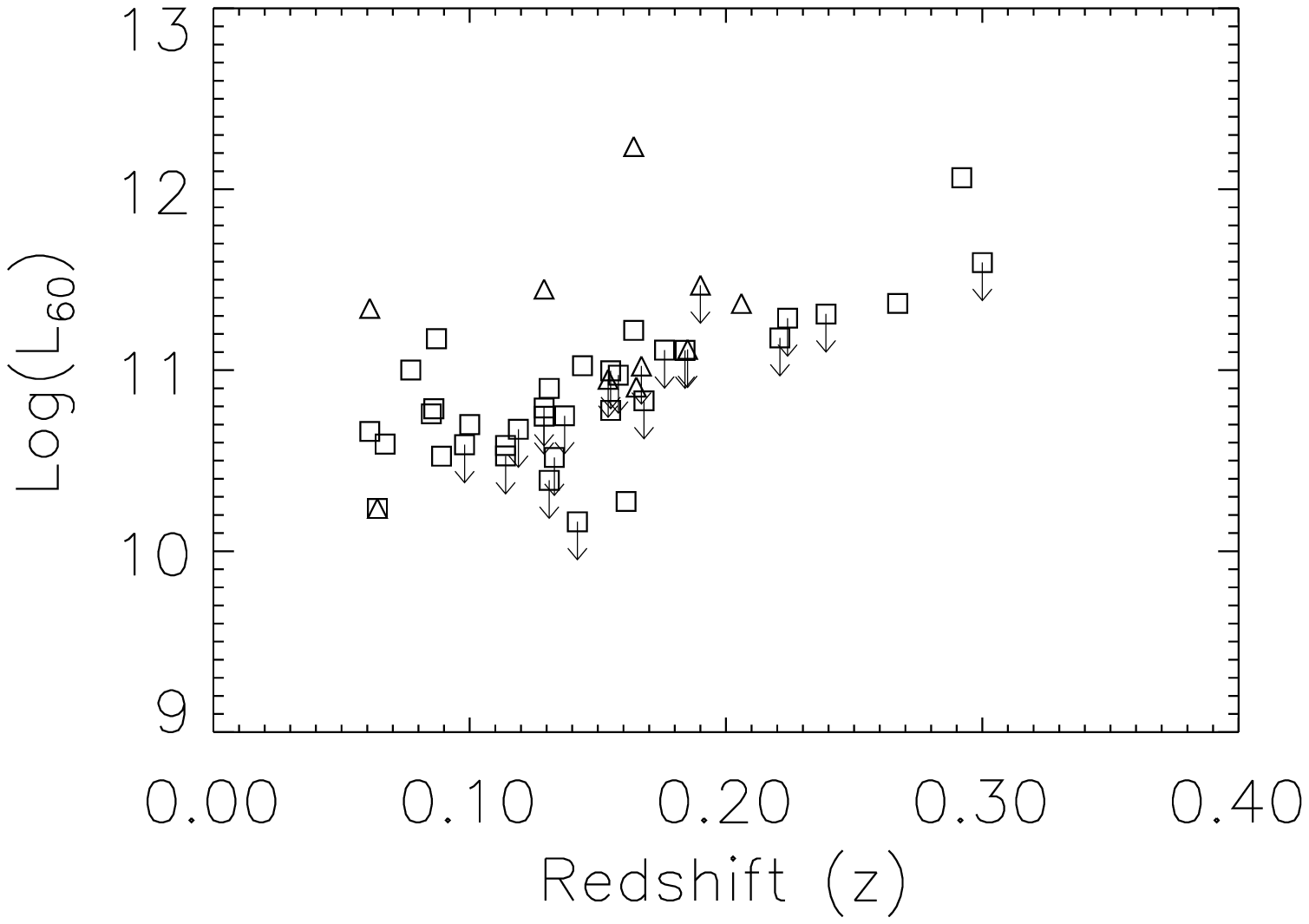,width=15cm}
\psfig{file=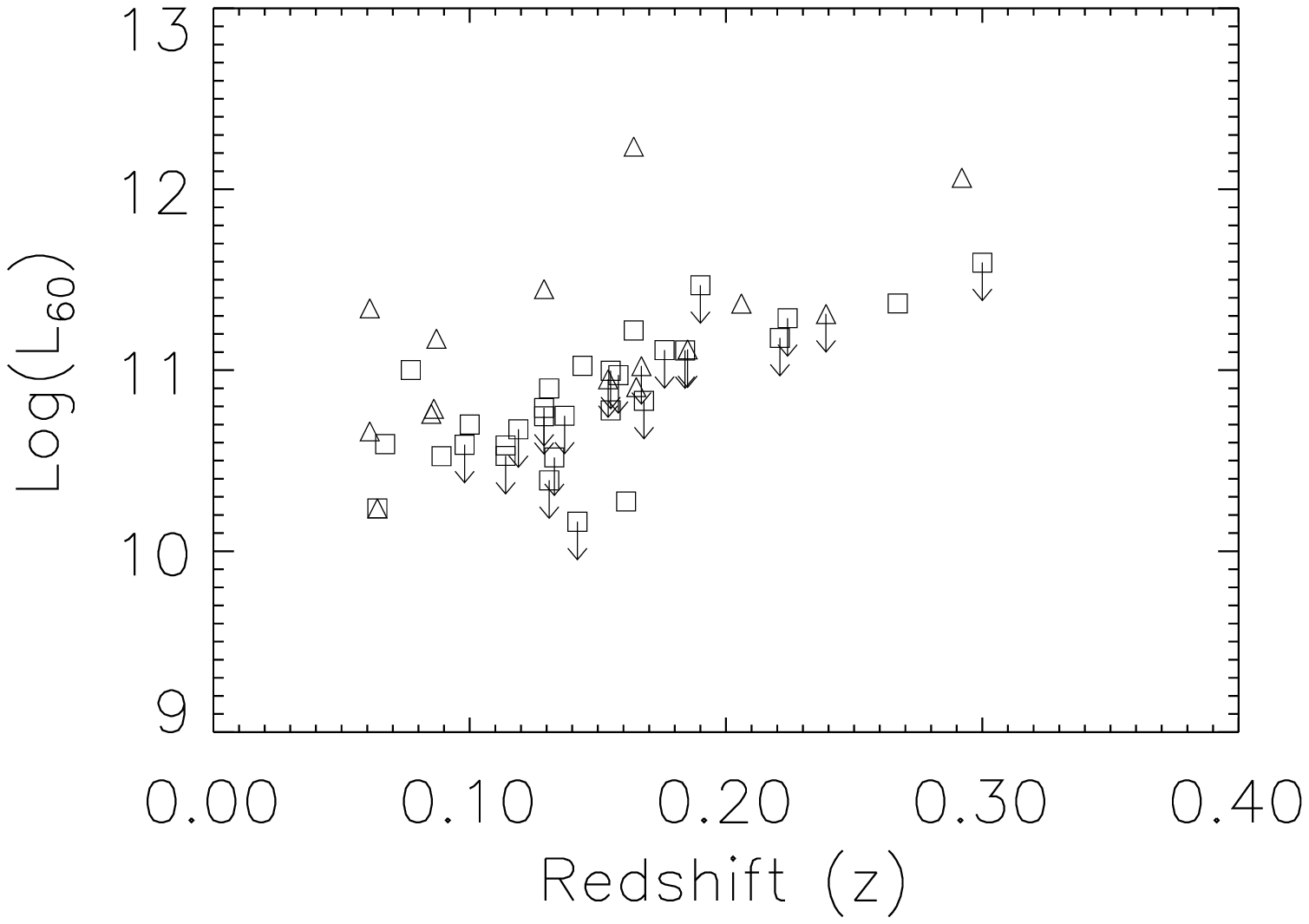,width=15cm}
\caption{FIR Properties of Quasars in Disturbed and Undisturbed Hosts}

(Upper) Homogeneous morphological classification; (Lower) Inhomogeneous
morphological classification. Triangles are quasars in disturbed hosts,
squares are quasars with undisturbed hosts. A line descending from the
centre of the symbol indicates that the result is an upper limit. All
upper limits are 3$\sigma$, all detections are at $>3\sigma$.
\end{figure*}

The large number of upper-limits in this dataset indicates that a
full examination requires a Survival Analysis approach. These
techniques allow the proper handling of `censored' datasets -- ie. those
that contain many upper (or lower) limits. We apply a non-parametric
univariate survival analysis test to detect differences in the FIR
luminosities of the two classes of quasars.  This test makes no
assumptions about the underlying luminosity distributions of either
sample i.e. it is non-parametric. We are thus not reliant on any
particular models for the luminosity functions of these objects, which
is useful since many aspects of quasar and ULIRG luminosity functions
are still subject to debate.  The method is, however, still dependent
on the `censoring' pattern, i.e. the manner in which upper limits are
applied, in the compared samples. For example a bias might be
introduced if the disturbed-host quasars were observed more deeply
than the quiescent-host objects. However, no such bias is apparent in
the data.  We applied the Peto-Prentice statistical test to see if the
two distributions are different using the ASURV Rev 1.2 package
\cite{l92}, which implements the methods presented in Feigelson \&
Nelson \shortcite{f85}. This particular test was used since it is the
least sensitive to differences in censoring patters.

We thus test whether the 60$\mu$m luminosity distributions of
disturbed and undisturbed host quasars are different. 60$\mu$m
luminosities were used because more of the quasars were detected at
60$\mu$m than 100$\mu$m, minimising the effect of censoring
patterns. We find that the probability that the difference in
60$\mu$m luminosity distributions between quasars with disturbed and
undisturbed hosts could occur at random is 0.029 for the homogeneous
dataset, and 0.0035 for the inhomogeneous.  We thus have found a
difference between the two with 97.1\% confidence (99.65\% confidence
for the inhomogeneous set).

We also use the Kaplan-Meier estimator to derive values for the mean
60$\mu$m luminosity for each of the two classes of quasar. These were
found to be 0.9 $\pm 0.3 \times 10^{11} L_{\odot}$ for the quasars in
undisturbed hosts and 3 $\pm 1 \times 10^{11} L_{\odot}$ for the
quasars in disturbed hosts (0.5 $\pm 0.1 \times 10^{11}L_{\odot}$ and 3
$\pm 1 \times 10^{11}L_{\odot}$ respectively for the inhomogeneous
dataset).
While the small numbers of FIR luminous quasars
in this sample implies that these values should be treated with caution,
they nevertheless provide an interesting confirmation of the results
of the Peto-Prentice test.

We thus conclude that we have detected a significant difference in
60$\mu$m luminosity between the two classes, with the disturbed-host
quasars having luminosities 3--5 times greater than undisturbed-host
quasars. This is apparent in both the homogenously classified dataset
and is even more significant in the inhomogenously classified dataset.
This argues strongly that the effect we are seeing is real,
and that a full appraisal of the morphologies of these quasar hosts
with, for example, HST, would only strengthen the result further.

\subsection{FIR luminous galaxies containing quasars: A Coincidence?}

Quasars and luminous FIR galaxies are both very rare classes of
object. If their luminosity functions are independent, indicating that
the quasar activity is unrelated to the FIR luminosity,
then the joint luminosity function $\Phi(M_B,L_{FIR})$ will simply be

$$\Phi(M_B,L_{\rm FIR})=\frac{\Phi_{\rm QSO}(M_B)\Phi_{\rm
IRAS}(L_{\rm FIR})}
                                   {\Phi_{\rm IRAS}(L=0)}$$

For the absolute magnitude and FIR luminosity ranges of interest here,
$\Phi_{QSO}$ and $\Phi_{ULIRG}$ are roughly the same at
10$^{-6}$--10$^{-7}$ Mpc$^{-3}$ \cite{k98}. Assuming that the luminosity
functions are independent, we can then calculate the number of objects
expected in the survey volume with both high FIR luminosity and a
quasar in the absolute magnitude range of interest from
$\Phi(M_B,L_{FIR})$ multiplied by the volume of the survey. This
suggests we should expect $\sim$ 10$^{-2}$---$10^{-4}$ objects with
L$_{FIR}>10^{12}L_{\odot}$, when we actually find 4.  This
discrepancy, by a factor of at least 400, argues for a causal connection
between the FIR luminosity, the disturbed morphology and the quasar
activity.  We discuss these issues in the next section.

\section{Discussion}

\subsection{Completeness and Bias}

Whilst the PG quasar subsample we discuss above has complete imaging
data from McLeod \& Rieke (1994a,b), and plentiful additional data
from other authors, it seems likely that it does not meet the ideal
requirement that it be a complete sample of quasars. Recent results
from a number of workers (e.g. Goldschmidt et al. \shortcite{g92})
have suggested that significant incompleteness exists in the PG
survey, especially at the low redshifts of interest to host morphology
studies. We must thus be concerned that the undetermined causes of
this incompleteness might undermine the basis for the present
study. If absence of an object from the PG sample is correlated, for
example, with high FIR luminosity or disturbed host morphology, then
the results of the present study could be biased.

PG quasars were selected on the basis of UV excess, $U-B <-0.44$, and
point like appearance in the optical. One prime candidate for the
incompleteness is poor photometric accuracy
\cite{g92}. Photometric errors are random with respect to
the nature of the objects since they will come from differences in the
processing and analysis of each photographic plate on the
sky. Incompleteness coming from photometric errors is thus unlikely to
bias the present results.

Other effects, though, might be of greater concern. Examples include
contamination of the quasar light by host galaxy emission, diluting
the UV excess and dropping the quasar out of the PG selection; the presence
of extended emission causing the object to be classified as a galaxy and
not thus as a blue stellar object and candidate quasar; and finally,
reddenning of the quasar light by dust in a host galaxy, again eliminating
the UV excess.

If any of these effects favours the presence of FIR luminous quasars with
disturbed hosts being in the PG Quasar catalogue, or removes FIR luminous
quiescent host quasars, then the present result must be regarded as insecure.
A definitive answer to this question, though, must wait until we have complete
host imaging and FIR data for one of the new, complete, objectively selected
surveys of nearby quasars such as the Hamburg survey \cite{k97}.

\begin{itemize}

\item {\bf Colour Effects} The PG survey magnitudes were measured in a 5''
aperture by an automatic plate scanner. A substantial amount of host
galaxy light might thus be included, and thus the UV excess of the
quasar might easily be diluted. McLeod \& Rieke found that almost all
of the host galaxies in their survey were spiral like. These will thus
have $U-B$ colours between $-0.2$ and $0.5$, with most objects being at the
redder end. The host galaxies of the FIR luminous quasars are likely
to have similar properties to ULIRGs. Arp 220, the archetypal ULIRG,
has a $U-B = 0.33$. This is well within the range for normal
spirals, so ULIRG-like objects will be affected in
much the same way as objects with quiescent host galaxies.
We thus do not expect colour dilution by the host galaxy to
bias the present results.

\item {\bf Reddening} FIR luminous objects clearly contain a substantial
mass of dust which might also lead to reddening of the emitted
spectrum.  This effect can clearly lead to the elimination of a UV
excess and the loss of an object from the PG survey. This is perhaps
the most worrying possible bias for the present work. One could easily
imagine the presence of a population of reddened quasars that do not
have UV excesses but that do have strong FIR luminosities. These
objects would bias the present results if we make the unusual
assumption that they are preferentially in undisturbed hosts. This
idea can be observationally tested with the extensive flux limited
redshift surveys of IRAS galaxies. If there were a large population of
quiescent host, FIR luminous, reddened quasars, they would be found in
surveys such as PSCz (Saunders et al., in preparation) or FSCz
\cite{o95}.  This does not seem to be the case (Oliver, private
communication), since few unknown, FIR luminous quasars (i.e.  high
luminosity broad-line objects) appear in these surveys. It may well be
that a population of completely obscured, FIR luminous quasars exists,
analogous to Seyfert 2s and radio galaxies, but that is not relevant
to the present study which deals only with optically identified
broad-line objects. We thus conclude that reddening effects cannot
significantly bias the present results.

\item {\bf Extended Emission} While many objects in the PG quasar catalogue
do have extended nebulosities visible on the photographic plates
\cite{s83}, the survey was intended to select stellar images only. Thus
objects which show very clear extended emission will be rejected.  UV
excess stellar sources were selected by eye from photographic
plates. There are thus likely to be inconsistencies and
inhomogeneities in the definition of a stellar object, but this is
unlikely to be correlated with FIR properties. B band plate limits
ranged from $B=15.5$ to 16.67. Thus only fairly bright features would be
seen. The elimination of extended sources in this way could remove
quasars with bright host galaxies from the PG survey. The present
result could be biased by this effect if quiescent host galaxies are
preferentially excluded on the basis of their extended emission, and
some significant fraction of these had strong FIR emission.  There are
several counter arguments to this. An undisturbed quasar host
galaxy will have the central engine lying at its own peak in surface
brightness while a disturbed host quasar is more likely to have peaks
in surface brightness offset from the central engine. A quasar in a
disturbed host is thus more likely to be classified as extended than
an undisturbed host. The absence of a significant number of previously
unknown FIR-bright broad line objects in complete IRAS surveys, as
discussed above, also argues against this idea. However, an
alternative is that there is a population of disturbed-host quasars
with {\em low} FIR luminosity that have been systematically overlooked
as candidate PG objects.  These could reduce the mean FIR luminosity
of the disturbed-host class and reduce the significance of the result
found in Section 3. For this to happen, the emission would have to
appear above the plate limit and be sufficiently well--separated from
the quasar nucleus for it to be seen as extended.  This is clearly
easier at lower redshifts, so we might expect there to be a deficit of
lower redshift disturbed-host quasars in the PG survey.  However, the
disturbed host quasars appear over the whole of the limited redshift
range in the present study, which argues against this
possibility. This scenario, though, remains a possible alternative to
the association of high FIR luminosity and disturbed quasar host
morphology.
\end{itemize}

\subsection{Far Infrared Emission in Quasar Hosts}

We have found that there is a significant difference between the FIR
luminosities of quasars in disturbed and undisturbed host
galaxies. The mean luminosity of the disturbed-host quasars is
several times greater than that of the quiescent-host quasars. These
results argue that the FIR emission from quasars may have more to do
with the host galaxy than with the central engine, and that it may
have a similar origin to the strong FIR emission seen in ULIRGs.  For
many ULIRGs the dominant power source in the FIR seems to be a
starburst \cite{g98}.  Various authors have already used different
lines of evidence to reach similar conclusions. Rowan-Robinson
\shortcite{rr95} argues that the infrared emission of a sample of
IRAS--detected PG quasars is coming from two components: hot dust in
the narrow--line region dominating the 3--30$\mu$m emission, and a
cooler, starburst powered component at 30--100$\mu$m. This latter
component would dominate the FIR luminosity that appears to be
unusually strong in the disturbed--host quasars.  Sopp \& Alexander
\shortcite{sa91}, meanwhile, have used the results of radio
observations to argue that there is a common origin for the radio and
FIR emission in starbursts and radio-quiet quasars. This is based on
the result that radio-quiet quasars lie on the same radio-FIR
correlation as starbursts and other star-forming galaxies.  In the
light of these results and our own, it seems likely that the FIR
emission from quasars may have more to tell us about the host galaxy
than the central engine itself. The results from quasar surveys with
ISO (eg. Wilkes et al. \shortcite{w97}) should provide more
information on this issue, and should be interpreted with host
properties in mind.  The situation is somewhat confused, however, by
the recent results of Blundell et al. (1996, 1998)
who argue, on the basis of small scale nuclear jets
seen in radio quiet quasars with the VLBA, for a black-hole origin for
the radio emission. Quite how these objects would then appear on the
same radio-FIR relation as starbursts is unclear.

\subsection{The Nature of Quasar Evolution}

We find that the numbers of FIR luminous quasars are
substantially higher than would be expected if the quasar and high FIR
luminosity were occurring independently. Together with the result that
high FIR luminosity is associated with a disturbed host morphology,
this looks suspiciously like the ULIRG-into-quasar scenario originally
proposed by Sanders et al. \shortcite{s88a}. In this scheme quasars are
triggered by galaxy mergers. The merger causes material in the
interacting galaxies to fall towards the centre of the merged system.
This appears to trigger a massive starburst, powering much, if not
all, of the prodigious FIR luminosity seen in ULIRGs.  Our new result
might then confirm the earlier idea that some FIR-luminous galaxies
also harbour quasars which are formed or re--awakened by the
merger. As these systems evolve, the quasar gradually becomes visible,
while signs of the merger decline, both in terms of enhanced FIR
luminosity and disturbed host morphology. Quasars that lie in clearly
disturbed systems would thus be at an earlier stage of evolution than
the bulk of the population, and have stronger FIR luminosity.

However, it remains to be shown that the disturbed--host quasars are
at an early stage of their development, since we have merely shown
that their host galaxies show all the signs of a recent ULIRG-like
interaction event.  To do this one needs to examine the properties of
the central engine itself as revealed, for example, by its emission
and absorption line properties, or by its X-ray properties (Perry,
private communication).  If FIR luminous or disturbed--host quasars
were found to deviate from the bulk of the quasar population in any
central-engine properties, we would have not only a plausible reason
to suggest that they were at a different stage in their evolution, but
also an indication of the physics of that evolution. In this context
it is interesting to note work by Lipari \shortcite{li94} on
extreme FeII emitting AGN, many of which are FIR luminous.

If mergers do indeed trigger quasar activity, via a ULIRG stage, then
we may have an interesting new perspective on the so--called
quasar epoch at z=2--3 (see eg. Dunlop \&
Peacock \shortcite{d90} and Shaver et al. \shortcite{sh96}). Current
galaxy formation theories suggest that galaxies formed by the
merging of smaller sub-units (eg. Kauffman
\shortcite{k95}). Observations of the Hubble Deep Field
(eg. Clements \& Couch \shortcite{c96c}, Abraham et
al. \shortcite{a96}) show that many high redshift, candidate primeval
galaxies are indeed very disturbed, possibly merging, systems. If high
redshift mergers trigger ULIRG-like activity, as they appear to in the
nearby universe, and ULIRGs lead in some cases to quasars, then we
could speculate that the epoch of rapid galaxy sub-unit merging might
also be the epoch of strongest quasar activity. The z=2--3
quasar--epoch might thus be indicating the time when most galaxies
underwent their major merging phase.

\subsection{Confirmation of this result}

While this result seems fairly secure, for the reasons expounded
above, the current state of knowledge of quasar hosts and FIR emission
is not so good that we can be absolutely sure of our
conclusions. Before we can say that, a number of caveats must be
investigated.

Firstly, we must improve our knowledge of the local quasar host
population.  For the present study we have either had to use limited
IR morphological data, or had to bring together results from disparate
studies at a range of wavelengths and with differing depths and
angular resolutions. This is not an ideal situation. Far better would
be a consistent survey of all the low redshift PG quasars dealt with
in this paper reaching faint fluxes (eg. reaching surface brightness
levels of R=26 mag arcsec$^{-1}$ as in McLure et al. \shortcite{m98})
with good angular resolution (0.5 arcsec or better). This might be
achieved with a large HST survey, though it is interesting to note
that only ten of the 45 PG quasars discussed here have been observed
since its launch. Alternatively, large ground-based telescopes
equipped with adaptive optics might achieve this goal.

Secondly, we must be sure that there is no significant population of
disturbed-host non-FIR luminous quasars which, as discussed above,
might have been missed by the PG quasar survey, and produce biases in
the present study. This could be achieved by mounting a host galaxy
imaging survey of recent objectively selected quasar surveys eg. the
Hamburg survey \cite{k97} with similar parameters as the study
suggested above for PG quasars.

Finally, our knowledge of the FIR emission of quasars must also be improved.
Some steps towards this have been made by ISO \cite{w97}, but
these samples are still quite small and will not have reached the necessary
sensitivity in many cases. Instead we must look to SIRTF and SOFIA as
our best hopes for obtaining the necessary FIR observations at high
enough sensitivities. The new generation of submillimetre bolometer
arrays (eg. SCUBA) might also have a role to play by providing
submillimetre rather than FIR fluxes.

\section{Conclusions}

We have compared the FIR properties of quasars with disturbed and
undisturbed host galaxies in a redshift and absolute optical magnitude
limited subsample of the PG quasar survey. We find that those objects
with disturbed hosts have a significant tendency to having greater FIR
luminosities. We also find that FIR luminous quasars occur in greater
numbers than would be expected if the quasar and high FIR luminosities
were independent.  This result is consistent with the view that ULIRGs
and quasars are related phenomena. However, there are several reasons
why these results might be incomplete or biased, including the
morphological classifications which are based on an inhomogeneous set
of observations, and the use of the PG survey itself, which may be
biased and incomplete. Specifically, if the PG survey is biased away
from selecting non-FIR luminous quasars with interacting hosts, the
present results would have to be revised. Improved quasar host
morphologies, from homogeneous deep high resolution surveys, and
better FIR data would allow us to place the present result on a firmer
footing.  Such morphological studies are possible with HST, or
ground--based adaptive optics systems, while improved FIR data will
have to await SIRTF or SOFIA.
\\~\\
{\bf Acknowledgements}
\\~\\
I would like to thank Steve Serjeant, Amanda Baker, Loretta Dunne, Kim
McLeod, Judith Perry, Pippa Goldschmidt, Steve Eales and Seb Oliver
for useful discussions. Special thanks to Loretta Dunne and Amanda
Baker for doing the independent morphological classifications. This
work was supported by the `Surveys with the Infrared Space
Observatory' network set up by the European Commission under contract
ERB FMRX-CT96-0068 of its TMR programme and by a PPARC postdoctoral
grant.

\end{document}